\documentclass[aps,pre,onecolumn,superscriptaddress,floatfix]{revtex4-1}

\usepackage{graphicx}
\usepackage[latin1]{inputenc}
\usepackage{amsfonts}
\usepackage{amssymb}
\usepackage{amsmath,amssymb}
\usepackage{bm}
\usepackage{cancel}
\usepackage{cleveref}
\usepackage{subfigure} 
\usepackage{bbold}
\usepackage{caption}
\usepackage{color}
\newtheorem{remark}{Remark}

\newtheorem{definition}{Definition}

\newcommand{\nc}{\newcommand}

\nc{\NN}{\mathbbm{N}} %
\nc{\ZZ}{\mathbbm{Z}} %
\nc{\QQ}{\mathbbm{Q}} %
\nc{\II}{\mathbbm{I}} %
\nc{\RR}{\mathbbm{R}} %
\nc{\CC}{\mathbbm{C}} %

\newcommand{\qt}{\vspace{-0.2cm} \begin{quotation} \noindent }
\newcommand{\eqt}{\end{quotation}}

\begin{document}


\author{L.~L. Bonilla}
\affiliation{G. Mill\'an Institute, Department of Mathematics, Universidad Carlos III de Madrid, Avenida de la Universidad 30,  Legan\'es 28911, Spain, e-mail: bonilla@ing.uc3m.es}
 \author{M. Carretero}
 \affiliation{G. Mill\'an Institute, Department of Mathematics, Universidad Carlos III de Madrid, Avenida de la Universidad 30,  Legan\'es 28911, Spain, e-mail: manuel.carretero@uc3m.es}
\author{F. Terragni}
\affiliation{G. Mill\'an Institute, Department of Mathematics, Universidad Carlos III de Madrid, Avenida de la Universidad 30,  Legan\'es 28911, Spain, e-mail: filippo.terragni@uc3m.es}

  \title{Integrodifference master equation describing actively growing blood vessels in angiogenesis}

\begin{abstract}
We study a system of particles in a two-dimensional geometry that move according to  a reinforced random walk with transition probabilities dependent on the solutions of reaction-diffusion equations for the underlying fields. A birth process and a history-dependent killing process are also considered. This system models tumor-induced angiogenesis, the process of formation of blood vessels induced by a growth factor released by a tumor. Particles represent vessel tip cells, whose trajectories constitute the growing vessel network. New vessels appear and may fuse with existing ones during their evolution. Thus, the system is
  described by tracking the density of active tips, calculated as an ensemble average over many realizations of the stochastic process. Such density satisfies a novel discrete master equation with source and sink terms. The sink term is proportional to a space-dependent and suitably fitted killing coefficient. Results are illustrated studying two influential angiogenesis models.
  \end{abstract}



\maketitle


\section{Introduction}

In the present paper, we investigate the density of a set of particles undergoing a reinforced random walk, branching and killing processes in two space dimensions. Branching and killing are point processes but the latter occurs only when a particle intersects the trajectory of another particle or collides with it. These systems have received a lot of attention as they model angiogenesis, the growth of blood vessel networks that are important in organ development, wound healing, and in many pathologies including cancer \cite{AC98,pla04}. However, most of the theoretical work consists of computational models or mathematical models that are solved numerically and directly compared with experiments, with little mathematical elaboration \cite{sci13,hec15}. The goal of the present paper is to derive a master equation for the density of the active particles representing the angiogenic network and validate it by comparing its solutions to those obtained by direct simulation of the underlying stochastic process.

The growth of blood vessels out of a primary vessel or angiogenesis is a complex multiscale process responsible for organ growth and regeneration, tissue repair, wound healing, and many other natural operations in living beings \cite{car05,CT05,GG05,fruttiger,CJ11}. Angiogenesis is triggered by lack of oxygen (hypoxia) experienced by cells in some tissue. Such cells secrete growth factors that diffuse and reach a nearby primary blood vessel. In response, the vessel wall opens and issues endothelial cells that move towards the hypoxic region. The cells at the tips of the advancing network are highly motile, do not proliferate, and react to gradients of the growth factors released at the hypoxic region and to mechanical cues. Following them, other endothelial cells proliferate and build the blood vessels. Thus, capillaries bring blood, oxygen, and nutrients to the hypoxic region. Many models assume that the angiogenic network is made out of the moving vessel tips (considered as point particles) and their trajectories. New particles are created by a branching process, while particles that intersect the trajectory of another particle are destroyed, as the vessels they represent merge with a preexisting blood vessel, in a process called anastomosis. Once blood and oxygen have reached the hypoxic region, secretion of growth factors stops, anti-angiogenic substances may be released and a regular vessel network is put in place, after pruning capillaries with insufficient blood flow.

In healthy circumstances, angiogenic and anti-angiogenic activities balance. Imbalance may result in many diseases including cancer \cite{fol71,car05,fol06}. In fact, after a tumor installed in a tissue reaches some two millimeters size, it needs additional nutrients and oxygen to continue growing. Its hypoxic cells secrete growth factors that induce angiogenesis. Unlike normal cells, cancerous ones continue issuing growth factors and attracting blood vessels, which also supply them with a handy transportation system to reach other organs in the body. In angiogenesis, events happening at cellular and subcellular scales unchain endothelial cell motion and proliferation, and build millimeter size blood sprouts and networks thereof \cite{CT05,GG05,fruttiger,CJ11}. Models range from very simple to extraordinarily complex and often try to illuminate some particular mechanism; see the reviews \cite{sci13,hec15}. Realistic microscopic models involve postulating mechanisms and a large number of parameters that cannot be directly estimated from experiments, but they often yield qualitative predictions that can be tested. An important challenge is to extract mesoscopic and macroscopic descriptions of angiogenesis from the diverse microscopic models.

Early models were based on reaction-diffusion equations (RDEs) for growth factor densities, endothelial cell densities, etc. \cite{lio77,cha93,cha95} and they could not describe the growing blood vessel network. In a seminal work, Anderson and Chaplain \cite{AC98} derived a reinforced random walk for growing blood vessels from the RDE describing the density of endothelial cells. Then, they supplemented the random walk with branching and anastomosis processes. Unfortunately, the transition probabilities obtained from a finite difference discretization of a PDE may become negative for some parameter ranges and therefore the resulting random walk with additional branching and killing rules can be inconsistent. On the other hand, Plank and Sleeman introduced related reinforced random walks with explicitly non-negative transition probabilities \cite{pla04}. By connecting stochastic processes to RDEs, it is possible to ascertain how densities representing substances important for cell adhesion, anti-angiogenic substances, and other continuum fields influence a growing angiogenic network \cite{cha06}. In early reinforced random walk models \cite{AC98,pla04,cha06,ste06}, solutions of the RDEs are not affected by the reinforced random walk describing the advancing network. Thus, the angiogenic network may be seen as postprocessing the RDEs. Models using stochastic differential equations to describe blood vessel extension (instead of random walks) couple density or flux of vessel tips to RDEs for growth factors or fibronectin \cite{cap09,bon14,bon16,bon16pre,ter16,bon17mbe,bon17}. Therefore, concentrations of the latter are also random variables. This is also the case of multiscale models of vascular tumor growth that account for the angiogenesis part of the overall process by means of reinforced random walks \cite{mac09}.

The reinforced random walk setting \cite{spitzer} has been widely adopted to study biological systems \cite{othmer97}. However,
to our knowledge, literature on reinforced random walk models for angiogenesis reports only numerical simulations of vessel networks \cite{AC98,pla04,cha06,mac09,hec15}. In this work, in order to quantify the process of angiogenic network growth induced by a tumor, we define a density of active blood vessel tips, as similarly done in models based on stochastic differential equations \cite{cap09,bon14,ter16,bon17mbe}.
Then, in the framework of two widely studied models \cite{AC98,pla04},
we derive a deterministic, discrete master equation for the density of active tips calculated from ensemble averages over replicas of the underlying stochastic process.
Note that a density of active tips describes an advancing angiogenic network \cite{cap09,bon14,ter16,bon17mbe} and is quite different from a density of endothelial cells or from a probability density of extending vessel tips \cite{AC98,pla04,cha06,ste06,mac09,hec15}.
Here, the deterministic description consists of a discrete-time integrodifference equation for the mentioned density coupled to discretized RDEs for the growth factor and relevant continuum fields.
The most important, novel term characterizes vessel fusion and is nonlocal in time. Its space-dependent coefficient must be estimated by
comparison with numerical simulations of the involved stochastic process.
Note that seeking deterministic equations for the density of active blood vessel tips may help developing qualitative and quantitative analyses of more complex models of angiogenesis, such as those reviewed in \cite{hec15}.

The paper is organized as follows. In section~\ref{sec:2}, we describe the reinforced random walk and point processes modeling angiogenesis. In
section~\ref{sec:3}, we derive a discrete master equation for the density of active tips and compare its solution with the ensemble-averaged tip density calculated from the underlying stochastic process. Finally, section~\ref{sec:4}
contains our conclusions and future research directions.


\section{Stochastic model} \label{sec:2}

In order to model tumor-induced angiogenesis, we first follow \cite{AC98} and consider a system of nondimensional coupled RDEs for the density of endothelial cells (ECs), $n(t,\mathbf{x})$, a growth factor (GF) concentration, $C(t,\mathbf{x})$, and fibronectin (FN) concentration, $f(t,\mathbf{x})$, namely
\begin{eqnarray}
\frac{\partial n}{\partial t} & = & D\,\Delta n - \nabla \cdot \left( \frac{\chi}{1 + \alpha C}\,n\,\nabla C \right) - \nabla \cdot \left( \rho\,n\,\nabla f \right)\!, \label{eq1} \\
\frac{\partial C}{\partial t} & = & - \eta\,n\,C, \label{eq2} \\
\frac{\partial f}{\partial t} & = & \beta\,n - \gamma\,n\,f. \label{eq3}
\end{eqnarray}
In eq.~\eqref{eq1}, the flux of ECs consists of diffusion, $-D\nabla n$, chemotaxis proportional to the GF gradient, $\chi n \nabla C/(1+\alpha C)$, and haptotaxis proportional to the FN gradient, $\rho n \nabla f$. Diffusion arises from ECs Brownian motion, while chemotaxis indicates that ECs move in the direction of larger gradients of the GF released by the tumor. Haptotaxis is the ECs motion to larger adhesion gradients of FN, an adhesive macromolecule attached to the extra-cellular environment. The GF in eq.~\eqref{eq2} is simply degraded, whereas there is an uptake and production of FN as shown in eq.~\eqref{eq3}. Other processes important in different situations can be included by adding more terms and new RDEs \cite{cha06}. In these equations, time has been scaled by factor $\tau = L^2 / D_c$, where $L = 2$ mm is the characteristic length of the problem and $D_c = 2.9 \cdot 10^{-11}$ $\textrm{m}^2$/s is the GF diffusion coefficient. Values of the involved dimensionless parameters are given by $D = 0.00035$, $\chi = 0.38$, $\alpha = 0.60$,
$\rho = 0.20$, $\eta = 0.10$, $\beta = 0.05$, and $\gamma = 0.10$. Note that they
have been taken from \cite{AC98} except for $\rho = 0.20$, which is smaller here as to simulate a weaker haptotaxis.
\begin{figure}[h]
\begin{center}
\includegraphics[width=7.0cm]{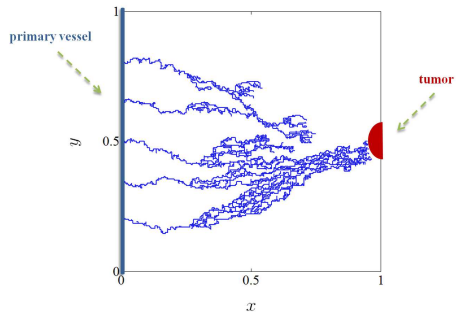}
\end{center}
\vskip-6mm
\caption{Sketch of the geometry for angiogenesis from a primary vessel to a circular tumor. \label{fig1}}
\end{figure}
We consider a simple square geometry (Figure~\ref{fig1}), in which a primary vessel is located at $x = 0$ and initially emits $5$ capillaries,
while a small circular tumor is centered at $(x,y) = (1.1,0.5)$ and acts as the GF source.
Equation~\eqref{eq1} is subject to zero-flux boundary conditions on $[0,1] \times [0,1]$, namely
\begin{equation}
\mathbf{\xi} \cdot \left( - D\,\nabla n + \frac{\chi}{1 + \alpha C}\,n\,\nabla C + \rho\,n\,\nabla f \right) = 0 \label{eqBCn}
\end{equation}
on the boundaries of the unit square (here, $\mathbf{\xi}$ is the outward unit normal vector).
Note that eq.~\eqref{eq2} lacks spatial derivatives, hence the initial GF concentration has to model the tumor, as we cannot impose a (perhaps more natural) boundary condition on the GF flux \cite{bon14}. Initial conditions for $n$, $C$, and $f$ are
\begin{equation}
n(0,x,y) \,=\, \sum_{j = 1}^3 e^{-x^2/0.0025 -(y-y_j)^2/0.0025}, \label{eq5}
\end{equation}
\begin{equation}
C(0,x,y) \,=\, e^{-(1.1-x)^2/0.45 -(0.5-y)^2/0.8}, \quad f(0,x,y) \,=\, 0.75\,e^{-x^2/0.45}, \label{eq6}
\end{equation}
where $y_1 = 0.2$, $y_2 = 0.5$, and $y_3 = 0.8$. \\

The stochastic process for (endothelial) vessel tips comprises a reinforced random walk, branching, and anastomosis, which is derived as follows.
Equations~\eqref{eq1}--\eqref{eq3} are spatially discretized on a $201 \times 201$ grid of the unit square by step $h = 0.005$. Spatial derivatives are approximated by second-order centered finite differences and the RDEs are integrated over time by the explicit Euler scheme with time step $\Delta t = 0.001$. After reorganizing terms, the fully discretized eq.~\eqref{eq1} for $n^q_{l,m} = n(q \Delta t,lh,mh)$ can be written as
\begin{equation}
n^{q+1}_{l,m} \,=\, n^{q}_{l,m} W_0 + n^{q}_{l+1,m} W_1 + n^{q}_{l-1,m} W_2 + n^{q}_{l,m+1} W_3 + n^{q}_{l,m-1} W_4, \label{eq8}
\end{equation}
with coefficients $W_0$, $W_1$, $W_2$, $W_3$, $W_4$ given by
\begin{eqnarray}
W_0 &=& 1-\frac{4D\Delta t}{h^2} + \frac{\alpha \Delta t \,\chi(C_{l,m}^q)}{4h^2(1+\alpha C_{l,m}^q)}\left[(C_{l+1,m}^q-C_{l-1,m}^q)^2+(C_{l,m+1}^q-C_{l,m-1}^q)^2\right] \nonumber \\
&-& \frac{\Delta t\,\chi(C_{l,m}^q)}{h^2}(C_{l+1,m}^q+C_{l-1,m}^q+C_{l,m+1}^q+C_{l,m-1}^q-4C_{l,m}^q) \nonumber \\
&-& \frac{\rho \Delta t}{h^2}(f_{l+1,m}^q+f_{l-1,m}^q+f_{l,m+1}^q+f_{l,m-1}^q-4f_{l,m}^q), \label{apeq1}
\end{eqnarray}
\begin{eqnarray}
\displaystyle W_1 &=& \frac{D \Delta t}{h^2}-\frac{\Delta t}{4h^2}\left[\chi(C_{l,m}^q)\,(C_{l+1,m}^q-C_{l-1,m}^q)+\rho(f_{l+1,m}^q-f_{l-1,m}^q)\right], \label{apeq2} \\[2mm]
\displaystyle W_2 &=& \frac{D \Delta t}{h^2}+\frac{\Delta t}{4h^2}\left[\chi(C_{l,m}^q)\,(C_{l+1,m}^q-C_{l-1,m}^q)+\rho(f_{l+1,m}^q-f_{l-1,m}^q)\right], \label{apeq3} \\[2mm]
\displaystyle W_3 &=& \frac{D \Delta t}{h^2}-\frac{\Delta t}{4h^2}\left[\chi(C_{l,m}^q)\,(C_{l,m+1}^q-C_{l,m-1}^q)+\rho(f_{l,m+1}^q-f_{l,m-1}^q)\right], \label{apeq4} \\[2mm]
\displaystyle W_4 &=& \frac{D \Delta t}{h^2}+\frac{\Delta t}{4h^2}\left[\chi(C_{l,m}^q)\,(C_{l,m+1}^q-C_{l,m-1}^q)+\rho(f_{l,m+1}^q-f_{l,m-1}^q)\right], \label{apeq5}
\end{eqnarray}
where $\chi(C) = \chi/(1+\alpha C)$ and $C^q_{l,m} = C(q \Delta t,lh,mh)$, $f^q_{l,m} = f(q \Delta t,lh,mh)$.
Equation~\eqref{eq8} resembles a master equation except that $W_i$ are not true transition probabilities because they are not normalized.
Let us then define transition probabilities for a reinforced random walk as
\begin{equation}
\mathbb{W}_i = \frac{W_i}{ \tilde{W}}, \quad \textrm{with } \,i = 0, \ldots, 4, \qquad \tilde{W} = \sum_{j = 0}^4 W_j. \label{eq9}
\end{equation}

\begin{definition}\label{rem1} (reinforced random walk) \,Given an initial configuration with $N_0 = 5$ equally spaced tips located at the grid nodes $(0,0.20)$, $(0,0.35)$, $(0,0.50)$, $(0,0.65)$, and $(0,0.80)$, their motion is determined by means of the transition probabilities in eq.~\eqref{eq9}. At each time step, a random number $U$ (from a uniform distribution between 0 and 1) is generated for each tip, which will stay at the current node if $U \in [0,\mathbb{W}_0]$, move to the left if $U \in (\mathbb{W}_0,\sum_{j = 0}^1 \mathbb{W}_j]$, move to the right if $U \in (\sum_{j = 0}^1 \mathbb{W}_j,\sum_{j = 0}^2 \mathbb{W}_j]$, move downwards if $U \in (\sum_{j = 0}^2 \mathbb{W}_j,\sum_{j = 0}^3 \mathbb{W}_j]$, or move upwards if $U \in (\sum_{j = 0}^3 \mathbb{W}_j,1]$. The transition probabilities are evaluated at the tip node,
movement is allowed only onto the nodes of the spatial grid and all tips reaching any of the four boundaries of the unit square are `deactivated'.
\end{definition}

\begin{definition}\label{rem5} (anastomosis) \,Whenever a tip meets an existing vessel (namely, the trajectory of another tip), it merges with the latter and is `deactivated'. Whenever a tip meets another tip, one of them is `deactivated' and the other one remains active.
\end{definition}

\begin{definition}\label{rem6} (branching) \,A new tip can branch out of any of the existing active tips with a certain probability, regardless of its age. This is different from \cite{AC98}, in which a minimum age is required for a tip to branch. Tip branching is simulated as follows. At a given time step and for an active tip at the node located at $\mathbf{x} = (x,y)$, a random number $U_b$ is extracted from a uniform distribution between $0$ and $0.005$. Then, a new tip branches out of that tip at the same node if $U_b \,\leq\, \phi(\mathbf{x})\,\Delta t$, where
\begin{eqnarray}
\phi(\mathbf{x}) &=& \left\{\begin{array}{lll} 0 & & \textrm{ if } \,\,\,0 \leq x \leq 0.25 \\[1mm]
1.5 & & \textrm{ if } \,\,\,0.25 < x \leq 0.45 \\[1mm]
2.5 & & \textrm{ if } \,\,\,0.45 < x \leq 0.65 \\[1mm]
3 & & \textrm{ if } \,\,\,0.65 < x \leq 0.85 \\[1mm]
4 & & \textrm{ if } \,\,\,0.85 < x \leq 1 \,.\end{array}\right. \label{eq10}
\end{eqnarray}
The upper bound of $U_b$ produces a nonzero probability that no tip branches out at a given time. Note that a newly branched tip will move according to Definition~\ref{rem1} and that branching onto an occupied node of the grid or any of the unit square boundaries is not allowed.
\end{definition}

\bigskip

\begin{figure}[h]
\begin{center}
\includegraphics[width=9.5cm]{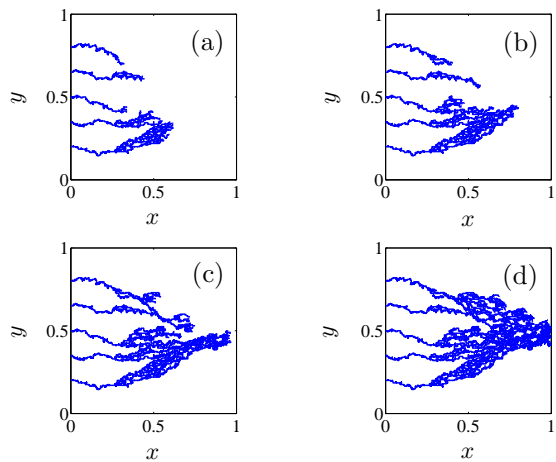}
\end{center}
\vskip-5mm
\caption{Growing vessel network on the underlying grid simulated with transition probabilities as in eqns.~\eqref{eq3.2}-\eqref{eq3.3} at (a) 5 days (35 active tips), (b) 6 days (36 active tips), (c) 7 days (52 active tips), (d) 8 days (38 active tips). \label{fig2}}
\end{figure}

In the same context, Plank and Sleeman proposed a different reinforced random walk that, in absence of branching and anastomosis, yields in the continuum limit the PDE in eq.~\eqref{eq1} for the probability density \cite{pla04}. The associated transition probabilities are
\begin{eqnarray}
\hat{\tau}_{l,m}^{H\pm} &=& \frac{4D}{h^2} \frac{\tau(w_{l\pm\frac{1}{2},m})}{\tau(w_{l+\frac{1}{2},m})+\tau(w_{l-\frac{1}{2},m})+\tau(w_{l,m+\frac{1}{2}})+\tau(w_{l,m-\frac{1}{2}})}, \label{eq3.2} \\[2mm]
\hat{\tau}_{l,m}^{V\pm} &=& \frac{4D}{h^2} \frac{\tau(w_{l,m\pm\frac{1}{2}})}{\tau(w_{l+\frac{1}{2},m})+\tau(w_{l-\frac{1}{2},m})+\tau(w_{l,m+\frac{1}{2}})+\tau(w_{l,m-\frac{1}{2}})}, \label{eq3.3}
\end{eqnarray}
where $w = (C,f)$ and $\tau(w) = \tau_1(C)\tau_2(f)$, with
\begin{eqnarray}
\tau_1(C) = (1+\alpha C)^{\frac{\chi}{\alpha D}}, \quad \tau_2(f) = e^{\rho f/D}. \label{eq41}
\end{eqnarray}
Note that in eqns.~\eqref{eq3.2}-\eqref{eq3.3} transitions are calculated by means of the nearest half-step neighbors
instead of the nearest neighbors \cite{othmer97}.
Tip motion, anastomosis, and branching can be treated as in Definitions~\ref{rem1}, \ref{rem5}, and \ref{rem6} above. Numerical simulations of the stochastic model with transition probabilities given by either eq.~\eqref{eq9} or eqns.~\eqref{eq3.2}-\eqref{eq3.3}
produce angiogenic networks that are quite similar.
A typical outcome is given in Figure~\ref{fig2}, which shows the growing of a vessel network on the underlying grid from the primary vessel
towards the tumor.
%


\section{Master equation} \label{sec:3}

In the configuration described in section~\ref{sec:2}, vessel tips tend to cluster in a relatively narrow region
due to the gradients of the GF and FN. Thus, many of them are killed by anastomosis during the evolution
and the number of active tips maintains rather small.
As a consequence, the stochastic model is not self-averaging and average quantities are not expected to resemble
those of a typical realization (i.e., a replica) of the process. Indeed,
different replicas of the stochastic process produce angiogenic networks (i.e., sets of the trajectories
of all actively moving vessel tips) that may look quite different.
Nevertheless, the tip density is preserved under the \textit{ensemble average} over a sufficiently large number of replicas
and thus a deterministic description can follow for ensemble-averaged quantities.
Let us then define the density of active tips at time $t$ as an ensemble average over realizations of the stochastic process.

\begin{definition} (ensemble-averaged tip density)
\,Consider all $N(t,\omega)$ active tips at time instant $t$ for a realization of the stochastic process labeled by $\omega$, with $\omega = 1,\ldots,\mathcal{N}$. Let $\mathbf{X}^i(t,\omega)$ be the node-location of the $i$-th tip at time instant $t$. Then, the ensemble-averaged tip density $P_{\mathcal N}(t,\mathbf{x})$ is defined as
\begin{equation}
P_{\mathcal N}(t,\mathbf{x}) \,=\, \frac{1}{\mathcal{N}}\sum_{\omega=1}^\mathcal{N} \sum_{i=1}^{N(t,\omega)}\delta_{\sigma}(\mathbf{x}-\mathbf{X}^i(t,\omega)), \quad \delta_{\sigma}(\mathbf{x}) \,=\, \frac{e^{-x^2/(\sigma_x)^2-y^2/(\sigma_y)^2}}{\pi\sigma_x\sigma_y}. \label{eq11}
\end{equation}
\end{definition}

\begin{figure}[h]
\begin{center}
\includegraphics[width=9.2cm]{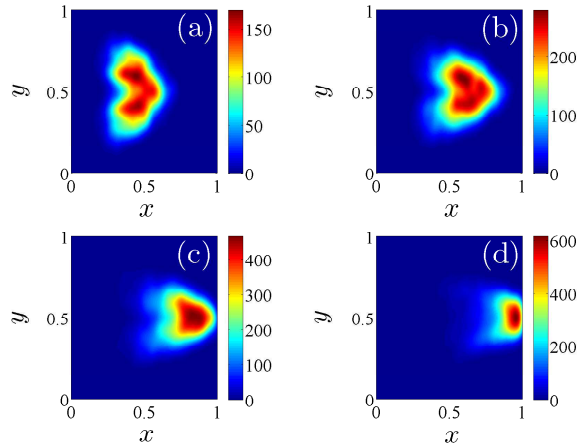}
\end{center}
\vskip-5mm
\caption{Ensemble-averaged (over $\mathcal N = 800$ replicas) stochastic tip density, when using transition probabilities as in eqns.~\eqref{eq3.2}-\eqref{eq3.3}, at (a) 5 days (19 active tips), (b) 6 days (32 active tips), (c) 7 days (41 active tips), (d) 8 days (29 active tips). \label{fig3}}
\end{figure}

Whenever it exists, the limit of $P_{\mathcal N}(t,\mathbf{x})$ as $\mathcal{N}\to\infty$ and $\sigma_x$, $\sigma_y\to0$ is the deterministic tip density $P(t,\mathbf{x})$. In our simulations, we use standard deviations equal to $\sigma_x = 0.03$ and $\sigma_y = 0.05$, being results fairly robust in connection with such values. Figure~\ref{fig3} shows time evolution of the ensemble-averaged tip density
with $\mathcal N = 800$, when using transition probabilities as in eqns.~\eqref{eq3.2}-\eqref{eq3.3} in the stochastic model.
Note that, once $P_{\mathcal N}(t,\mathbf{x})$ is computed, the ensemble-averaged number of active tips can also be recovered by integrating the density over the unit square and taking the integer part (this is used, for a better comparison with the master equation solution described below, after $P_{\mathcal N}(t,\mathbf{x})$ has reached the tumor).

\bigskip

We now derive a master equation for the density of active vessel tips corresponding to the stochastic process defined in section~\ref{sec:2}. The master equation should be discrete in time and contain the transition probabilities, from either eq.~\eqref{eq9} or eqns.~\eqref{eq3.2}-\eqref{eq3.3}. In addition, it should include terms related to branching and anastomosis. Let $P_{l,m}^{\,q} = P(q \Delta t,lh,mh)$ be the density of active tips at the grid node indexed by $(l,m)$ and time $t = q \Delta t$. Then, the discrete master equation associated with transition probabilities as in eq.~\eqref{eq9} is given by
\begin{eqnarray}
P_{l,m}^{\,q+1} &=& P_{l,m}^{\,q} \mathbb{W}_0 + P_{l+1,m}^{\,q} \mathbb{W}_1 + P_{l-1,m}^{\,q} \mathbb{W}_2 + P_{l,m+1}^{\,q} \mathbb{W}_3 + P_{l,m-1}^{\,q} \mathbb{W}_4 \nonumber \\ & + & \Delta t \,\,\phi_{l,m} \,P_{l,m}^{\,q} - \Delta t \,\,\Gamma_{l,m} \,P_{l,m}^{\,q} \sum_{k = 0}^q P_{l,m}^{\,k}. \label{meqac}
\end{eqnarray}
The last-but-one is a birth term modeling branching of new tips, where $\phi_{l,m}$ is the discrete counterpart of the branching function given in eq.~\eqref{eq10}. The last one is a nonlocal killing term corresponding to anastomosis, where $\Gamma_{l,m}$ is the discrete anastomosis function to be fitted by comparison with the ensemble-averaged results of the stochastic model, as discussed below. The anastomosis coefficient function depends on the spatial grid because the branching function does. Similar birth and death terms have been incorporated to Fokker-Planck-type equations in angiogenesis models that describe vessel extension by stochastic differential equations instead of random walks \cite{bon14,ter16,bon18}. It is worth remarking that the inclusion of both source terms in eq.~\eqref{meqac} is essential in order to properly describe an advancing vessel network and further analyze its behavior \cite{bon14,ter16,bon16,bon16pre,bon17,bon18}.

\begin{remark}\label{rem7} The transition probabilities $\mathbb{W}_i$, with $i = 0,\ldots,4$, depend on the solutions of eqns.~\eqref{eq1}--\eqref{eq3} calculated using the explicit Euler scheme as discussed in section~\ref{sec:2}. Thus, the evolution of the active vessel tips does not affect the density of ECs and the concentration of GF or FN.
\end{remark}

Integrodifference eq.~\eqref{meqac} has to be solved with absorbing boundary conditions, $P_{l,m}^q = 0$, at all nodes on the four boundaries of the unit square \cite{gardiner}. The initial density of active tips is set to
\begin{equation}
P_{l,m}^{\,0} \,=\, \frac{2}{\pi\sigma_x\sigma_y}\,e^{-x_l^2/(\sigma_x)^2} \sum_{i=1}^{N_0} e^{-(y_m-Y_i)^2/(\sigma_y)^2}, \label{initial_P_me}
\end{equation}
where $N_0 = 5$, $Y_1,\ldots,Y_5$ are the ordinates of the initial tips (see Definition~\ref{rem1}), and $(x_l,y_m) = (lh,mh)$.

\begin{remark}\label{rem8}
During time evolution, the number of active tips is set equal to $N_0 = 5$ as long as the $x$-location of the maximum of the tip density is smaller than $0.25$ (i.e., in absence of both branching and anastomosis, see eq.~\eqref{eq10} and eq.~\eqref{anast_fun} below). Otherwise, due to the structure of $P_{l,m}^{\,0}$ in eq.~\eqref{initial_P_me}, $N_0$ is the integer part of the sum $\sum_{l,m}P_{l,m}^{\,q}\,h^2$ over all grid nodes at each time instant $t = q \Delta t$.
\end{remark}

In the limit $h \rightarrow 0$ and $D/h^2 \rightarrow \infty$, the solution of eq.~\eqref{meqac} should approach the ensemble-averaged tip density provided by the stochastic model, namely $P_{\mathcal N}$ for sufficiently large $\mathcal N$ (see Figure~\ref{fig3}).

\begin{remark}\label{rem3}
Numerical simulations show that $\mathbb{W}_1$ in eq.~\eqref{eq9} may become negative during some time intervals. According to Definition~\ref{rem1} of the reinforced random walk, the tip cannot move to the left because $\mathbb{W}_0>\mathbb{W}_0+\mathbb{W}_1$ for these times. Thus motion to the left is impeded in favor of motion to the right and the resulting random motion toward the tumor is artificially accelerated.
\end{remark}

By construction, $\mathbb{W}_1$ depends on the values of $D$, $\chi$, $\alpha$, $\rho$ (together with the GF and FN derivatives), as seen in eqns.~\eqref{apeq1}--\eqref{eq9}.
As $\mathbb{W}_1$ turns out to be negative during various iterations of the stochastic simulations, the intervals over which we select probabilities are inconsistently defined, which  privileges advance toward the tumor. Hence, the resulting random motion will not be a true random walk, but it may still approximate well the solution of the PDE in eq.~\eqref{eq1}. Therefore, the transition rules in Definition~\ref{rem1} become inconsistent for some ranges of values of the mentioned quantities, which artificially privileges advance of the vessels toward the tumor. For the parameter values considered in this paper (which are the same as in \cite{AC98}), the discrete master equation solution
cannot be matched with the outcome of the reinforced random motion induced by eq.~\eqref{eq9} plus branching and anastomosis. \\

Now, in order to find a deterministic description that agrees with ensemble averages of the stochastic process discussed in section~\ref{sec:2},
let us take into account the (always positive) transition probabilities given by eqns.~\eqref{eq3.2}-\eqref{eq3.3}. The resulting discrete master equation is eq.~\eqref{meqac} with coefficients $\mathbb{W}_0,\ldots,\mathbb{W}_4$ given by
\begin{eqnarray}
&& \mathbb{W}_0 \,=\, 1 - (\hat{\tau}_{l,m}^{H+}+\hat{\tau}_{l,m}^{H-}+\hat{\tau}_{l,m}^{V+}+\hat{\tau}_{l,m}^{V-})\Delta t, \qquad
\mathbb{W}_1 \,=\, \hat{\tau}_{l+1,m}^{H-}\Delta t, \nonumber \\
&& \mathbb{W}_2 \,=\, \hat{\tau}_{l-1,m}^{H+}\Delta t, \qquad \mathbb{W}_3 \,=\, \hat{\tau}_{l,m+1}^{V-}\Delta t, \qquad
\mathbb{W}_4 \,=\, \hat{\tau}_{l,m-1}^{V+}\Delta t, \label{meqps}
\end{eqnarray}
and initial condition as in eq.~\eqref{initial_P_me}. Note that Remarks~\ref{rem7} and \ref{rem8} above still hold for the new discrete master equation~\eqref{meqac} based on eq.~\eqref{meqps}. Its resolution can be performed using the same spatial grid and $\Delta t = 0.001$ as
in the corresponding stochastic process.
Similarly to the work carried out in models based on stochastic differential equations \cite{ter16}, the discrete anastomosis function $\Gamma_{l,m}$ should be estimated so that the number of active tips calculated by ensemble averages of the stochastic model and by numerically solving the discrete master equation agree. Here, $\Gamma_{l,m}$ is the discrete counterpart of the function
\begin{eqnarray}
\Gamma(\mathbf{x}) &=& \left\{\begin{array}{lll} 0 & & \textrm{ if } \,\,\,0 \leq x \leq 0.25 \\[1mm]
0.0125 & & \textrm{ if } \,\,\,0.25 < x \leq 0.45 \\[1mm]
0.0210 & & \textrm{ if } \,\,\,0.45 < x \leq 0.65 \\[1mm]
0.0170 & & \textrm{ if } \,\,\,0.65 < x \leq 0.85 \\[1mm]
0.0160 & & \textrm{ if } \,\,\,0.85 < x \leq 1, \end{array}\right. \label{anast_fun}
\end{eqnarray}
which has been fitted in such a way that the maximum difference between the total number of active tips calculated via eqns.~\eqref{meqac}--\eqref{meqps} and by stochastic ensemble averages based on eqns.~\eqref{eq3.2}-\eqref{eq3.3} is equal to two for each time instant.
\begin{figure}[h]
\begin{center}
\includegraphics[width=8.0cm]{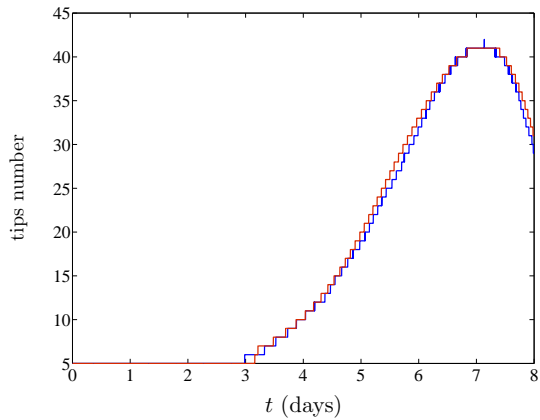}
\end{center}
\vskip-5mm
\caption{Time evolution of the total number of active tips as computed by the ensemble-averaged (over $\mathcal N = 800$ replicas) stochastic process based on eqns.~\eqref{eq3.2}-\eqref{eq3.3} (blue line) and the discrete master equation~\eqref{meqac} with eq.~\eqref{meqps} (red line). \label{fig4}}
\end{figure}

Figure~\ref{fig4} shows the time evolution of the total number of active tips according to the solution of the master equation and ensemble averages of stochastic simulations based on eqns. (\ref{eq3.2})-(\ref{eq3.3}). Both descriptions agree rather well. The number of active tips reaches a maximum and then decrease after the first ones arrive at the tumor. 

\begin{figure}[h]
\begin{center}
\includegraphics[width=9.2cm]{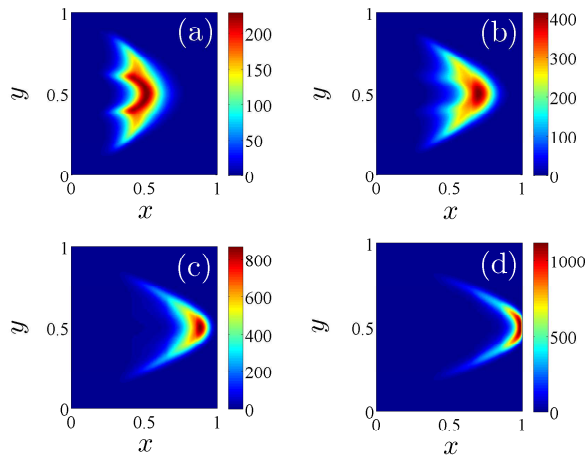}
\end{center}
\vskip-5mm
\caption{Tip density from the discrete master equation~\eqref{meqac} with eq.~\eqref{meqps}
at (a) 5 days (20 active tips), (b) 6 days (33 active tips), (c) 7 days (41 active tips), (d) 8 days (31 active tips). \label{fig5}}
\end{figure}

We have solved the discrete master equation~\eqref{meqac} with transition probabilities given by eq.~\eqref{meqps} for the anastomosis function given by eq.~\eqref{anast_fun}. The resulting density of active tips evolves in time as depicted in Figure~\ref{fig5}. This figure should be compared to the evolution of the active tip density calculated from ensemble averages and exhibited in Figure~\ref{fig3}. Note that the advance of the moving lump of active vessel tips is quite similar in both figures, although the shapes and sizes of the moving lumps are not the same. At the end of the formation stage, the lump obtained by ensemble averages of the random walk model is quite similar to that obtained by solving the master equation~\eqref{meqac}, as shown in Figures \ref{fig3}(a) and \ref{fig5}(a). Later on, the lump that solves \eqref{meqac}, shown in Figures \ref{fig5}(b)-(d), becomes narrower than that in Figures \ref{fig3}(b)-(d) while still keeping the same overall number of tips, as depicted in Figure \ref{fig4}. Therefore, the corresponding density inside the lump has to be larger for the solution of the master equation in Figure \ref{fig5} than for the ensemble averages of trajectories shown in Figure \ref{fig3}: the maximum density for the solution of the master equation may become almost twice the ensemble-averaged maximum density.

\begin{figure}[h]
\begin{center}
\includegraphics[width=7.0cm]{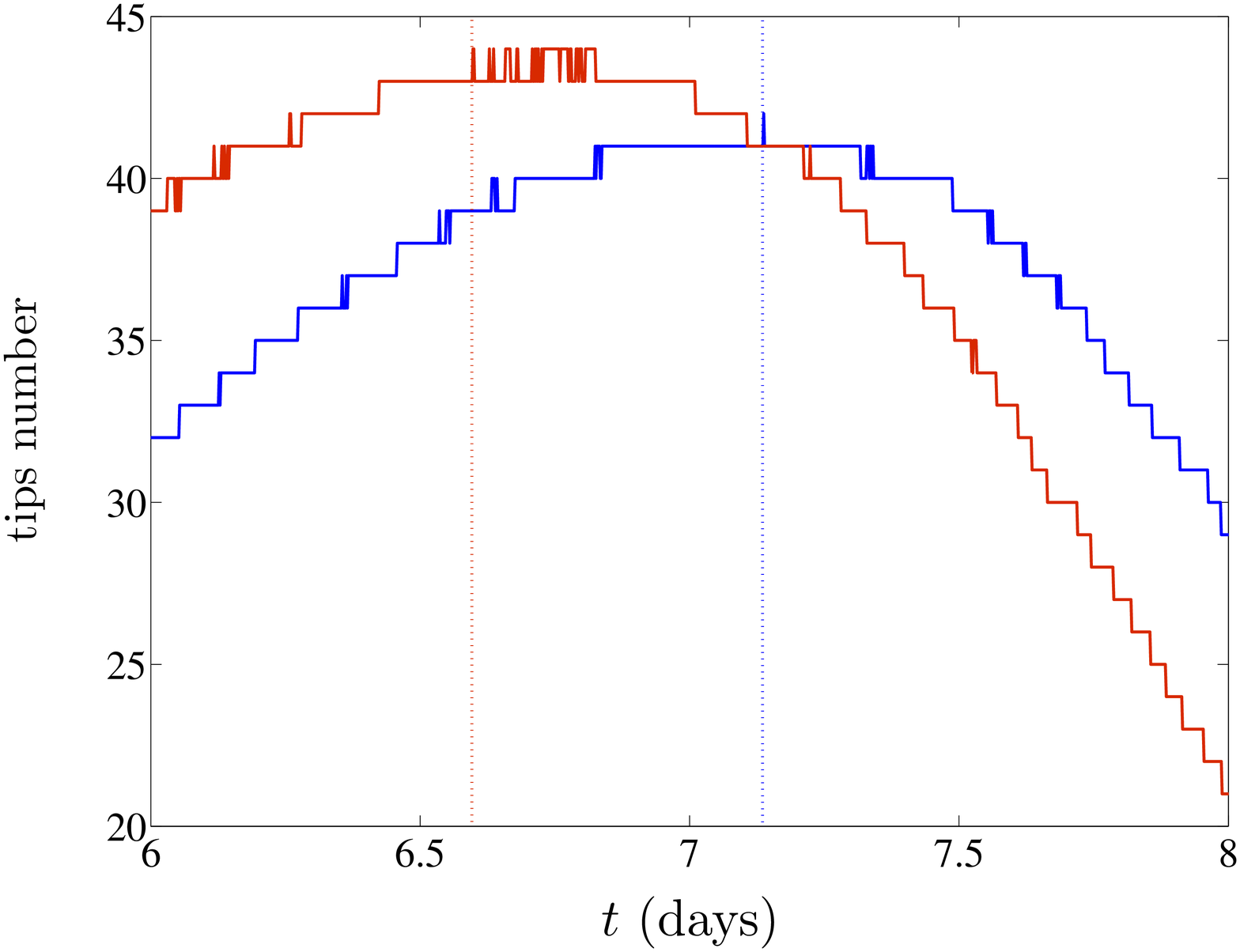}
\end{center}
\vskip-5mm
\caption{Comparison of the total number of active tips calculated from ensemble averages (over $\mathcal N = 800$ replicas) of the stochastic process based on eq.~\eqref{eq9} (red curve) and its counterpart based on eqns.~\eqref{eq3.2}-\eqref{eq3.3} (blue curve). \label{fig6}}
\end{figure}

Figure~\ref{fig6} compares the ensemble-averaged total number of active tips when using the transition probabilities given by
either eq.~\eqref{eq9} or eqns.~\eqref{eq3.2}-\eqref{eq3.3}.
This number decreases as the advancing network reaches the tumor. Thus, Figure~\ref{fig6} highlights that tips moving
according to eq.~\eqref{eq9}
arrive at the tumor before (about $0.5$ days earlier). Both random walks have the same continuum limit; additionally, branching and anastomosis are defined in the same way. Therefore, the faster evolution of the ensemble-averaged density based on eq.~\eqref{eq9} gives indirect evidence that the coefficient $\mathbb{W}_1$ in eq.~\eqref{eq9} becomes negative during the evolution of the angiogenic network. As explained in Remark \ref{rem3}, $\mathbb{W}_1<0$ means that the likelihood of moving to the right has artificially increased at the expense of the likelihood of moving to the left. This fact has been confirmed by inspection of the simulation data.
Observe that this stochastic description and the corresponding master equation would evolve over comparable time intervals (i.e., 7 days) only if the propagation velocity in the latter is enhanced by a factor $\approx\sqrt{2}$ (which is equivalent to consider in the master equation a characteristic timescale equal to $\hat{\tau} = \tau/\sqrt{2}$; see section~\ref{sec:2}). \\

The continuum limit of the discrete integrodifference master equation \eqref{meqac} based on eq.~\eqref{meqps} is
\begin{eqnarray}
\frac{\partial P}{\partial t}\!+\!\nabla\!\cdot\!\left(\frac{\chi\nabla C}{1+\alpha C}P\right)\!+\!\nabla\!\cdot\!(\rho \nabla f P)\!-\!D\Delta P\! \,=\, \!\phi(\mathbf{x})\, P\!-\!\Gamma(\mathbf{x})\, P\!\int_0^t\! P(s,\mathbf{x})ds. \label{pde}
\end{eqnarray}
This integrodifferential equation is similar to that for the marginal density of active tips derived in \cite{bon16,bon16pre} for a lattice-free model based on Langevin-Ito stochastic differential equations \cite{bon14,ter16} instead of reinforced random walks. A numerical finite difference scheme to solve eq.~\eqref{pde} is studied in detail in \cite{bon18}. There it is proved that the solutions of the scheme are positive, stable and converge to the solutions of the integrodifferential equation. The main differences between eq.~\eqref{pde} and the equations analyzed in those works are that $\phi(\mathbf{x})$ was a function of the GF concentration and $\Gamma(\mathbf{x})$ was a constant. These differences stem from the different definition of branching used in the model of \cite{bon16,bon16pre,bon18,ter16} (based on Langevin-Ito stochastic differential equations) and in the model considered here (based on \cite{AC98,pla04}). On the other hand,
numerical simulations of the stochastic process and solutions of the discrete master equation show that the density of active tips forms a moving lump after an initial stage. The profile of such lump at $y = 0.5$ is a solitary wave alike that found in tip cell models based on stochastic differential equations instead of reinforced random walks \cite{bon16,bon16pre,bon17}. It would be interesting to study the effects of discretization on the moving wave and whether the methodology used in \cite{bon16,bon16pre,bon17} is applicable to its motion despite the space-dependent coefficients in the source terms of eq.~\eqref{pde}.


\section{Conclusion and outlook} \label{sec:4}

In this paper, we have derived for the first time a discrete master equation for the density of active blood vessel tips starting from two well-known models of angiogenesis \cite{AC98,pla04}. These and other models based on reinforced random walks \cite{cha06,mac09} postulate that the densities of endothelial cells, growth factor, and appropriate substances solve a system of coupled reaction-diffusion equations. They propose reinforced random walks (either occurring on a lattice or being lattice-independent) whose continuum limit is the partial differential equation for the density of endothelial cells. Then, tip branching and vessel fusion are added to the random walks. We have shown that the evolution of the angiogenic network is described by the ensemble-averaged density of active tips. The latter solves a discrete master equation with birth (branching) and killing (anastomosis) terms, which are absent in all previous works. It is important to ensure that the terms corresponding to transition probabilities in this equation are non-negative, as done in \cite{pla04}. Otherwise, the solution of a master equation obtained from any numerical finite difference scheme (as done in \cite{AC98}) may evolve differently from ensemble averages of the active tips. In the continuum limit, this master equation becomes an integrodifferential equation similar to that describing the density of active tips in Langevin-Ito models \cite{bon14,ter16,bon16}. Rigorously establishing our derivations, and proving existence, uniqueness, and long-time stability results for the limiting master equation would be highly desirable, but clearly out of the scope of the present paper.

There are different interesting directions in which this work can be extended. The density of active tips satisfies an equation that is alike that of the density of endothelial cells except for the addition of source and sink terms. The latter is nonlocal in time. One interesting direction is obtaining a soliton description for the moving lump of Figures~\ref{fig3} and \ref{fig5}, as similarly done in \cite{bon16,bon16pre,bon17}. Finding equations for the fluctuations about the density of active tips (which is an average quantity) would be worthwhile also. The fluctuations would theoretically describe the statistics of the angiogenic network, a topic which is quite unexplored at the present time.

Another direction is extending the methodology of this paper to models of angiogenesis that probe cellular length scales. For example, cellular Potts models directly describe cells that change size and the extracellular matrix by Monte Carlo dynamics coupled to continuum fields for growth factors concentration, elastic fields, etc. (see reviews \cite{sci13,hec15} and articles \cite{bau07,van_oers,veg20}). It is challenging to deduce dynamics for the active vessel tips from the Monte Carlo cellular dynamics. Once this is done, it should be possible to derive a master equation with source terms for the active tip density. This could be more productive than deriving an equation for the density of endothelial cells with a given center of mass and a given perimeter, as in \cite{alb07}.

\acknowledgments
This work has been supported by the FEDER/Ministerio de Ciencia, Innovaci\'on y Universidades -- Agencia Estatal de Investigaci\'on grant MTM2017-84446-C2-2-R.


\end{document}